\documentclass[showpacs,superscriptaddress]{revtex4} 
\usepackage{epsfig,amsmath,amssymb,graphicx}
\usepackage{natbib}
\usepackage[english]{babel}

\begin{document}
\vspace{0mm}
\title{ THE TWO-PARAMETER DEBYE MODEL  } %
\author{Yu.M. Poluektov}
\email{yuripoluektov@kipt.kharkov.ua} %
\affiliation{National Science Center ``Kharkov Institute of Physics
and Technology'', Akhiezer Institute for Theoretical Physics, 61108
Kharkov, Ukraine} %
\affiliation{V.N.\,Karazin Kharkov National University, 61022
Kharkov, Ukraine }

\begin{abstract}
When analyzing thermodynamic and kinetic properties of crystals
whose anisotropy is not large and the considered effects do not
relate to the existence of singled-out directions in crystals, one
may use a more simple model of an isotropic medium with a good
accuracy, after having chosen its parameters in an optimal way.
Based on the quantum mechanical description it is shown that the
method of approximation of the moduli of elasticity of a crystal by
the model of an isotropic medium, proposed earlier in
\cite{Fedorov}, follows from the requirement of the minimal
difference between the free energies of a crystal and an
approximating isotropic medium. The two-parameter Debye model is
formulated, which, in contrast to the standard model where the
average velocity of phonons is introduced, takes into account the
existence in an isotropic medium of both longitudinal and transverse
phonons. The proposed model contains, except the Debye energy, an
additional dimensionless parameter and, consequently, the law of
corresponding states for the heat capacity being characteristic of
the standard model does not hold. With taking account of the two
phonon branches the structure of the density of phonon states proves
to be more complex as compared to the standard model and has a
singularity that resembles Van Hove singularities in real crystals.
As an example, an application of the two-parameter Debye theory to
such crystals of the cubic system as tungsten, copper, lead is
considered. It is shown that the calculation of the low-temperature
heat capacity of these crystals by means of the approximated moduli
of elasticity within the framework of the two-parameter model leads
to a considerably better agreement with experiment than in the case
of the standard Debye model.
\newline%
\end{abstract}
\maketitle

\section{Introduction}\vspace{-0mm} 
The elastic properties of an isotropic medium are characterized by
two elastic moduli, for example, the Lame coefficients $\lambda$ and
$\mu$ \cite{LL}. A crystal is an anisotropic medium, and therefore
its elastic properties differ from those of an isotropic body and
are characterized by a large number of parameters. If the degree of
the anisotropy is sufficiently small, then, obviously, the
properties of a crystal will slightly differ from the properties of
an isotropic medium and the analysis of the properties of a crystal
can be replaced by a more simple analysis of the properties of a
model isotropic medium, which parameters are selected in an optimal
way. In \cite{Fedorov} it is proposed to find the elastic moduli of
an approximating isotropic body from the condition of minimum of the
quantity
\begin{equation} \label{01}
\begin{array}{l}
\displaystyle{%
  G\equiv \big(\lambda_{iklm} - \lambda_{iklm}^{(0)}\big)^2,
}
\end{array}
\end{equation}
where $\lambda_{iklm}$ is the elastic moduli tensor of a crystal, and %
\begin{equation} \label{02}
\begin{array}{l}
\displaystyle{%
  \lambda_{iklm}^{(0)}=\lambda\delta_{ik}\delta_{lm} + \mu\big(\delta_{il}\delta_{km}+\delta_{im}\delta_{kl} \big),
}
\end{array}
\end{equation}
is the elastic moduli tensor of an isotropic medium. The extremum
conditions $\partial G\big/\partial\lambda=\partial
G\big/\partial\mu=0$ give a relation between the approximating Lame
coefficients and the invariants of the elastic moduli tensor of a
crystal:
\begin{equation} \label{03}
\begin{array}{l}
\displaystyle{%
  \lambda=\frac{1}{15}\big(2\lambda_{iikk}-\lambda_{ikik}\big),\qquad
  \mu=\frac{1}{30}\big(3\lambda_{ikik}-\lambda_{iikk}\big).
}
\end{array}
\end{equation}
This approximation is quite reasonable for a qualitative, and in
many cases quantitative, analysis of the integral properties of a
crystal and those effects that are not related to the existence of
distinguished directions in crystals and can be described in the
approximation of an isotropic medium. In \cite{Fedorov} such an
approach was developed for the theory of elastic waves in crystals.
A similar approach can be used when nonlinear effects are taken into
account in crystals, for example, to simplify the calculations of
the matrix elements of the interaction between phonons in crystals
of any system \cite{AAKh,ABKh}. In this paper we restrict ourselves
to the framework of the linear theory of elasticity.

Condition (\ref{01}) is convenient in that it leads to a system of
linear algebraic equations. However, in principle, other criteria
for the closeness of the elastic moduli tensors of a real and a
model media are possible, which contain, for example, the difference
modulus of a higher degree. It is natural to require that the
approximating elastic moduli are chosen so that the free energies of
real and model media would be as close as possible.

The purpose of this work is to show on the basis of quantum
consideration that the elastic moduli of an isotropic model medium,
obtained from the condition of the extremum of the quantity $G$
(\ref{01}), do indeed lead to the free energy being as close as
possible to the free energy of a crystal.

The well-known Debye model \cite{LL2} is widely used to describe the
thermodynamic properties of solids in the approximation of an
isotropic medium. In the standard Debye model, which is formulated
for an isotropic medium, a further simplification is made consisting
in the fact that an isotropic medium, instead of two parameters, is
characterized by a single parameter -- the average phonon velocity
and its corresponding energy. However, such a simplification is not
necessary and it seems natural to formulate a theory for an
isotropic medium with two elastic moduli and, accordingly, with two
types of phonons -- longitudinal and transverse. Obviously, such a
theory cannot be less accurate than the Debye model in the standard
formulation \cite{LL2} and, possibly, will allow us to describe some
more subtle effects. In addition, using the approximation of the
elastic moduli (\ref{03}), it becomes possible to apply the model to
the calculation of crystals of various systems. Thus, another goal
of this work is to formulate the Debye model which takes into
account the existence of longitudinal and transverse phonons in an
isotropic medium. As an example, we consider the application of the
proposed two-parameter Debye theory to crystals of the cubic system.
The approximating elastic moduli are found for tungsten, copper, and
lead. The calculation of their low-temperature heat capacities using
the calculated moduli shows that the two-parameter model describes
the thermodynamic properties with a much better approximation than
the standard Debye model.

\section{OPTIMAL APPROXIMATION OF ELASTIC PROPERTIES OF A CRYSTAL
\newline BY THE MODEL OF AN ISOTROPIC MEDIA}\vspace{-0mm} %
The density of the Hamiltonian of a crystal, as an elastic medium,
is given by the expression
\begin{equation} \label{04}
\begin{array}{l}
\displaystyle{%
  {\rm H}({\bf r})=\frac{\pi_a({\bf r})^2}{2\rho}+
  \frac{1}{2}\lambda_{aibj}u_{ai}({\bf r})u_{bj}({\bf r}), %
}
\end{array}
\end{equation}
where the deformation tensor in the linear approximation has the
form
\begin{equation} \label{05}
\begin{array}{l}
\displaystyle{%
  u_{ij}=\frac{1}{2}\big(\nabla_ju_i+\nabla_iu_j\big),
}
\end{array}
\end{equation}
$u_{i}({\bf r})$ is the displacement vector, $\pi_{a}({\bf r})=\rho \dot{u}_{a}({\bf r})$ %
is the canonical momentum, $\rho$ is the density, $\lambda_{aibj}$
is the elastic moduli tensor of a crystal. In (\ref{04}) and in the
following, the rule of summation over repeated indices is used.

Let us try to approximate the Hamiltonian (\ref{04}), which contains
the elastic moduli of a real crystal, by a model isotropic medium,
choosing its Hamiltonian in the form
\begin{equation} \label{06}
\begin{array}{l}
\displaystyle{%
  {\rm H}_S({\bf r})=\frac{\pi_a({\bf r})^2}{2\rho}+
  \frac{1}{2}\tilde{\lambda}_{aibj}u_{ai}({\bf r})u_{bj}({\bf r})+\varepsilon_0, %
}
\end{array}
\end{equation}
where
\begin{equation} \label{07}
\begin{array}{l}
\displaystyle{%
  \tilde{\lambda}_{aibj}=\tilde{\lambda}\delta_{ai}\delta_{bj} + \tilde{\mu}\big(\delta_{ab}\delta_{ij}+\delta_{aj}\delta_{bi} \big),
}
\end{array}
\end{equation}
and the Lame coefficients $\tilde{\lambda}$, $\tilde{\mu}$ will be
considered as parameters determined from the condition of the best
approximation of the crystal Hamiltonian (\ref{04}) by the
Hamiltonian of an isotropic medium (\ref{06}), which will be
introduced below. Here and in what follows the elastic moduli of an
approximating isotropic medium will be denoted by the tilde sign at
the top. The Hamiltonian (\ref{06}) contains the energy density
$\varepsilon_0$, which is caused by the fact that replacing the
exact elasticity tensor $\lambda_{aibj}$ with the approximate
isotropic one $\tilde{\lambda}_{aibj}$ can lead, generally speaking,
to a change in the ground undeformed state of a crystal as well. The
total initial Hamiltonian $H=\int {\rm H}({\bf r})d{\bf r}$ can be
represented in the form $H=H_S+H_C$, where the total approximating
Hamiltonian is singled out:
\begin{equation} \label{08}
\begin{array}{l}
\displaystyle{%
 H_S=\int\!\left[\frac{\pi_a({\bf r})^2}{2\rho}+
  \frac{1}{2}\tilde{\lambda}_{aibj}u_{ai}({\bf r})u_{bj}({\bf r})\right]\!\!d{\bf r}+V\varepsilon_0, %
}
\end{array}
\end{equation}
and the correlation Hamiltonian
\begin{equation} \label{09}
\begin{array}{l}
\displaystyle{%
 H_C=\frac{1}{2}\int\!\left[\big(\lambda_{aibj}-\tilde{\lambda}_{aibj}\big)\nabla_iu_a\nabla_ju_b\right]\!d{\bf r}-V\varepsilon_0 %
}
\end{array}
\end{equation}
characterizes the difference between the exact and approximating
Hamiltonians.

In the quantum description, which will be used, the deformation
vector $u_a({\bf r})$ and the canonical momentum $\pi_a({\bf r})$
should be considered as operators for which the known commutation
relations hold
\begin{equation} \label{10}
\begin{array}{l}
\displaystyle{%
 \pi_a({\bf r})u_b({\bf r}')-u_b({\bf r}')\pi_a({\bf r})=-i\hbar\delta_{ab}\delta({\bf r}-{\bf r}'),  %
}\vspace{2mm}\\ %
\displaystyle{%
 u_a({\bf r})u_b({\bf r}')-u_b({\bf r}')u_a({\bf r})=0, \qquad \pi_a({\bf r})\pi_b({\bf r}')-\pi_b({\bf r}')\pi_a({\bf r})=0.   %
}\vspace{2mm}\\ %
\end{array}
\end{equation}
Let us use the expansion of the field operators
\begin{equation} \label{11}
\begin{array}{l}
\displaystyle{%
  u_a({\bf r})=\frac{1}{\sqrt{V}}\sum_{k,\alpha}\sqrt{\frac{\hbar}{2\rho\omega({\bf k},\alpha)}}\,e_a({\bf k},\alpha)\big(b_{k\alpha}+b^+_{-k\alpha}\big)e^{i{\bf k}{\bf r}},  %
}\vspace{2mm}\\ %
\displaystyle{%
  \pi_a({\bf r})=-\frac{i}{\sqrt{V}}\sum_{k,\alpha}\sqrt{\frac{\rho\hbar\omega({\bf k},\alpha)}{2}}\,e_a({\bf k},\alpha)\big(b_{k\alpha}-b^+_{-k\alpha}\big)e^{i{\bf k}{\bf r}},  %
}%
\end{array}
\end{equation}
where ${\bf e}({\bf k},\alpha)$ are the complex polarization vectors
$(\alpha=1,2,3)$, such that ${\bf e}(-{\bf k},\alpha)={\bf e}^*({\bf k},\alpha)$, %
for which the conditions of orthogonality and completeness hold
\begin{equation} \label{12}
\begin{array}{l}
\displaystyle{%
  {\bf e}({\bf k},\alpha){\bf e}^*({\bf k},\alpha')=\delta_{\alpha\alpha'}, \qquad  %
  \sum_\alpha e_i^*({\bf k},\alpha)e_j({\bf k},\alpha)=\delta_{ij}.
}
\end{array}
\end{equation}
The creation $b_{k\alpha}^+$ and annihilation $b_{k\alpha}$
operators of phonons obey the usual commutation relations: %
$\big[b_{k\alpha},b_{k'\alpha'}^+\big]=\delta_{kk'}\delta_{\alpha\alpha'}$, %
$\big[b_{k\alpha},b_{k'\alpha'}\big]=\big[b_{k\alpha}^+,b_{k'\alpha'}^+\big]=0$. %
When describing an isotropic medium, it is convenient to choose one
of the polarization vectors directed along the wave vector:
${\bf e}({\bf k},3)=i\big({\bf k}/k\big)=i{\bf n}$. %
The other two polarization vectors with $\alpha=1,2$ lie in a plane
perpendicular to the wave vector. With this in mind, the
approximating Hamiltonian (\ref{08}) can be written in the diagonal
form
\begin{equation} \label{13}
\begin{array}{l}
\displaystyle{%
 H_S=\sum_k\left(2\hbar\omega_{kt}b_{kt}^+b_{kt}+\hbar\omega_{kl}b_{kl}^+b_{kl}\right)+ %
 \sum_k\!\left(\hbar\omega_{kt}+\frac{1}{2}\hbar\omega_{kl}\right)+V\varepsilon_0. %
}
\end{array}
\end{equation}
Here phonons with the transverse polarization ($\alpha=1,2$) are
denoted by the single index $t$, and phonons with the longitudinal
polarization ($\alpha=3$) are denoted by the index $l$. It is also
taken into account that the transverse phonons have two degrees of
freedom. As a result of reducing the Hamiltonian to the diagonal
form (\ref{13}), we find the phonon dispersion laws
$\omega_{kt}=c_tk$, $\omega_{kl}=c_lk$, where the velocities of the
transverse and longitudinal phonons are given by the well-known
expressions \cite{LL}:
\begin{equation} \label{14}
\begin{array}{l}
\displaystyle{%
 c_t=\sqrt{\frac{\tilde{\mu}}{\rho}}, \qquad c_l=\sqrt{\frac{\tilde{\lambda}+2\tilde{\mu}}{\rho}}. %
}
\end{array}
\end{equation}

In the following we will calculate the average quantities by means
of the statistical operator
\begin{equation} \label{15}
\begin{array}{l}
\displaystyle{%
 \hat{\rho}=\exp\beta\big(F-H_S\big),
}
\end{array}
\end{equation}
where $\beta=1/T$ is the inverse temperature. The normalization
condition ${\rm Sp}\hat{\rho}=1$ leads to the formula for the free
energy in the self-consistent field model
\begin{equation} \label{16}
\begin{array}{l}
\displaystyle{%
 F=V\varepsilon_0 + \sum_k\!\left(\hbar\omega_{kt}+\frac{1}{2}\hbar\omega_{kl}\right) + %
 2T\sum_k\ln\!\big(1-e^{-\beta\hbar\omega_{kt}}\big) + T\sum_k\ln\!\big(1-e^{-\beta\hbar\omega_{kl}}\big). %
}
\end{array}
\end{equation}
The energy of the undeformed state $\varepsilon_0$ in such a model
is found from the condition of equality of the averages for the
exact and approximating Hamiltonians $\langle H\rangle=\langle H_S\rangle$, %
or, equivalently, from the condition $\langle H_C\rangle=0$. This
gives
\begin{equation} \label{17}
\begin{array}{l}
\displaystyle{%
 \varepsilon_0=\frac{1}{2V}\int\!\left[\big(\lambda_{aibj}-\tilde{\lambda}_{aibj}\big)\Big\langle\nabla_iu_a\nabla_ju_b\Big\rangle\right]\!d{\bf r}.%
}
\end{array}
\end{equation}
Calculating the average in (\ref{17}) by means of the statistical
operator (\ref{15}), we arrive at the following formula
\begin{equation} \label{18}
\begin{array}{l}
\displaystyle{%
 \varepsilon_0=\frac{\hbar}{4\rho V}\big(\lambda_{aibj}-\tilde{\lambda}_{aibj}\big)\sum_k k_ik_j\Big[A_{kt}\delta_{ab}+\big(A_{kl}-A_{kt}\big)n_an_b\Big],%
}
\end{array}
\end{equation}
where the quantities
\begin{equation} \label{19}
\begin{array}{l}
\displaystyle{%
 A_{kt}=\frac{1+2f_{kt}}{\omega_{kt}}, \qquad  A_{kl}=\frac{1+2f_{kl}}{\omega_{kl}} %
}
\end{array}
\end{equation}
are expressed through the distribution functions of the longitudinal
$f_{kl}=\big[\exp\beta\hbar\omega_{kl}\,-1\big]^{-1}$ %
and transverse $f_{kt}=\big[\exp\beta\hbar\omega_{kt}\,-1\big]^{-1}$
phonons. The sums over wave vectors entering into (\ref{18}), after
passing from summation to integration and integrating over angles,
can be represented in the form
\begin{equation} \label{20}
\begin{array}{l}
\displaystyle{%
  \sum_k k_ik_jA_{kt}=\frac{V}{6\pi^2}\delta_{ij}I_t, \quad%
  \sum_k k_ik_jn_an_b\big(A_{kl}-A_{kt}\big)=\frac{V}{30\pi^2}\big(\delta_{ij}\delta_{ab}+\delta_{ia}\delta_{jb}+\delta_{ib}\delta_{ja}\big)\big( I_t-I_l\big), %
}
\end{array}
\end{equation}
where
\begin{equation} \label{21}
\begin{array}{l}
\displaystyle{%
  I_t=\int_0^{k_D}\!\!dk\,k^4\frac{\big(1+2f_{kt}\big)}{\omega_{kt}}=\frac{k_D^4}{4c_t}\Phi\!\left(\frac{\Theta_t}{T}\right), \quad %
  I_l=\int_0^{k_D}\!\!dk\,k^4\frac{\big(1+2f_{kl}\big)}{\omega_{kl}}=\frac{k_D^4}{4c_l}\Phi\!\left(\frac{\Theta_l}{T}\right). %
}
\end{array}
\end{equation}
Here $\Theta_t=\hbar c_tk_D,\, \Theta_l=\hbar c_lk_D$ are the Debye
energies defined through the velocities of the transverse and
longitudinal phonons, $k_D=\big(6\pi^2N/V\big)^{1/3}$ is the Debye
wave number, $N$ is the particle number. In (\ref{21}) there is
defined the function
\begin{equation} \label{22}
\begin{array}{l}
\displaystyle{%
  \Phi(x)=1+\big(8/3x\big)D(x), \quad D(x)=\frac{3}{x^3}\int_0^{x}\!\!\frac{z^3dz}{e^z-1}. %
}
\end{array}
\end{equation}
As a result, we find
\begin{equation} \label{23}
\begin{array}{l}
\displaystyle{%
 \varepsilon_0=\frac{3NT}{16V}\left[\left(\frac{\lambda_l}{x_l}-x_l\right)\!\Phi\big(x_l\big)+2\left(\frac{\lambda_t}{x_t}-x_t\right)\!\Phi\big(x_t\big)\right], %
}
\end{array}
\end{equation}
where for brevity we introduce the following notation:
\begin{equation} \label{24}
\begin{array}{l}
\displaystyle{%
  x_l\equiv\frac{\Theta_l}{T}=\frac{\hbar c_lk_D}{T}, \quad x_t\equiv\frac{\Theta_t}{T}=\frac{\hbar c_tk_D}{T}, \quad %
  \lambda_l\equiv\frac{\hbar^2k_D^2}{15\rho T^2}\big(2\lambda_B+\lambda_A\big),\quad  %
  \lambda_t\equiv\frac{\hbar^2k_D^2}{30\rho T^2}\big(3\lambda_B-\lambda_A\big).  %
}
\end{array}
\end{equation}
Here $\lambda_A\equiv\lambda_{iiaa}$,
$\lambda_B\equiv\lambda_{aiai}$ are the invariants of the crystal
elastic moduli tensor. As a result, taking into account (\ref{16}),
we obtain the free energy as a function of temperature, volume,
number of particles and two parameters $x_l, x_t$:
\begin{equation} \label{25}
\begin{array}{l}
\displaystyle{%
 F=\frac{3NT}{16}\left[\Phi\big(x_l\big)\!\left(\frac{\lambda_l}{x_l}-x_l\right)+2\Phi\big(x_t\big)\!\left(\frac{\lambda_t}{x_t}-x_t\right)\right]+\frac{3NT}{8}\big(x_l+2x_t\big)\,+ %
}\vspace{2mm}\\ %
\displaystyle{\hspace{5mm}%
  +\frac{NT}{3}\Big[3\ln\!\big(1-e^{-x_l}\big)-D\big(x_l\big)\Big] + \frac{2NT}{3}\Big[3\ln\!\big(1-e^{-x_t}\big)-D\big(x_t\big)\Big]. %
}%
\end{array}
\end{equation}
For arbitrary values of the parameters $x_l, x_t$, the free energy
(\ref{25}) describes a system in a state of incomplete thermodynamic
equilibrium. The equilibrium values of these parameters and,
consequently, the optimal values of the elastic coefficients of a
model isotropic medium $\tilde{\lambda}$ and $\tilde{\mu}$ should be
found from the conditions of the extremum of the free energy
(\ref{25}):
\begin{equation} \label{26}
\begin{array}{l}
\displaystyle{%
  \frac{\partial F}{\partial x_l}=0, \qquad  \frac{\partial F}{\partial x_t}=0. %
}
\end{array}
\end{equation}
As a result, we obtain the equations:
\begin{equation} \label{27}
\begin{array}{l}
\displaystyle{%
  \left(\frac{\lambda_\alpha}{x_\alpha^2}-1\right)\!\Big[x_\alpha\Phi'\big(x_\alpha\big)-\Phi\big(x_\alpha\big)\Big]=0, %
}
\end{array}
\end{equation}
where $\alpha=l,t$. The function in the second brackets does not
vanish in the field of its definition, so that the relation
$x_\alpha^2=\lambda_\alpha$ should hold, or in more detail:
\begin{equation} \label{28}
\begin{array}{l}
\displaystyle{%
  c_l^2=\frac{1}{15\rho}\big(2\lambda_B+\lambda_A\big), \qquad  %
  c_t^2=\frac{1}{30\rho}\big(3\lambda_B-\lambda_A\big).  %
}
\end{array}
\end{equation}
Hence, according to (\ref{14}), formulas for the Lame coefficients
of an isotropic medium that model a crystal have the form
\begin{equation} \label{29}
\begin{array}{l}
\displaystyle{%
  \tilde{\lambda}=\frac{1}{15}\big(2\lambda_A-\lambda_B\big), \qquad  %
  \tilde{\mu}=\frac{1}{30}\big(3\lambda_B-\lambda_A\big).  %
}
\end{array}
\end{equation}
These formulas coincide with the formulas (\ref{03}) obtained on the
basis of another criterion introduced in \cite{Fedorov}. As was also
shown here, such approximation proves to be valid at an arbitrary
temperature. Here are the approximating formulas for the bulk
modulus $\tilde{K}$, the Young modulus $\tilde{E}$ and the Poisson
ratio $\tilde{\sigma}$:
\begin{equation} \label{30}
\begin{array}{l}
\displaystyle{%
  \tilde{K}=\frac{1}{9}\lambda_A, \qquad  \tilde{E}=\frac{\lambda_A\big(3\lambda_B-\lambda_A\big)}{3\big(3\lambda_A+\lambda_B\big)}, \qquad  %
  \tilde{\sigma}=\frac{2\lambda_A-\lambda_B}{3\lambda_A+\lambda_B}.  %
}
\end{array}
\end{equation}
Note that the bulk modulus of the approximating continuous medium is
determined by the single invariant of a crystal $\lambda_A\equiv\lambda_{iiaa}$. %

\section{THE TWO-PARAMETER DEBYE MODEL}\vspace{-0mm} %
The Debye model \cite{LL2} describes the thermodynamic properties of
an isotropic elastic medium. The elastic properties of such a medium
are characterized by two elastic moduli, which can be chosen, for
example, as the Lame coefficients $\lambda$ and $\mu$ or any two
other moduli (\ref{30}). Accordingly, there are two sound modes
corresponding to the longitudinal and transverse vibrations and the
two velocities $c_l$ of $c_t$ of propagation of such waves
(\ref{14}). In the standard Debye theory \cite{LL2} an additional
simplification is used: instead of two velocities, the average
velocity of sound vibrations $c_D$ is introduced, which is defined
by the relation:
\begin{equation} \label{31}
\begin{array}{l}
\displaystyle{%
  \frac{1}{c_D^3}=\frac{2}{c_t^3}+\frac{1}{c_l^3}=\rho^{3/2}\left[\frac{2}{\mu^{3/2}}+\frac{1}{\big(\lambda+2\mu\big)^{3/2}}\right]. %
}
\end{array}
\end{equation}
A consequence of this approximation is that an isotropic elastic
medium is characterized by a single parameter $\Theta_D=\hbar c_D
k_D$ -- the Debye energy, and the heat capacity proves to be a
function of only the ratio $\Theta_D\big/T$. The definition
(\ref{31}) means that the inverse cubes of velocities are actually
averaged. However, one can define the average velocity in another
way, for example, by the relation
\begin{equation} \label{32}
\begin{array}{l}
\displaystyle{%
  c_0^2=\frac{1}{3}\big(2c_t^2+c_l^2\big)=\frac{\big(\lambda+4\mu\big)}{3\rho},  %
}
\end{array}
\end{equation}
which arises in an approach based on the description of interacting
phonons in the self-consistent field model
\cite{Poluektov,Poluektov2}. The definition (\ref{32}) seems more
natural, since for $\mu\rightarrow 0$ and $c_t\rightarrow 0$ from
(\ref{31}) it follows that $c_D\rightarrow 0$ as well, while the
definition (\ref{32}) gives in this case a finite value of the
average velocity $c_0$. The ratio of the average velocities defined
by the formulas (\ref{31}) and (\ref{32}) depends on the ratio of
the Lame coefficients
\begin{equation} \label{33}
\begin{array}{l}
\displaystyle{%
  \frac{c_0^2}{c_D^2}=\frac{1}{3^{5/3}}\big(4+\lambda/\mu\big)\!\left[2+\frac{1}{\big(2+\lambda/\mu\big)^{3/2}}\right]^{2/3}, %
}
\end{array}
\end{equation}
where $\lambda/\mu>-2/3$\, \cite{LL}. The dependence of the
velocities ratio $c_0\big/c_D$ on the ratio $\lambda/\mu$ is shown
in Fig.\,1. As the ratio $\lambda/\mu$ increases, the difference in
the average velocities determined by the formulas (\ref{31}) and
(\ref{32}) increases. The question may arise which averaging should
be considered correct. If we consider the average velocity and the
Debye energy as phenomenological parameters, then this question,
apparently, is not fundamental. At the same time, it is obvious that
the transition from the use of two parameters characterizing an
elastic medium to a single parameter is not necessary, and it is
more natural to construct a theory with taking into account both
phonon velocities. In doing so the complication of the theory proves
to be insignificant, and the question as to the way of introducing
the average velocity does not arise at all. It can be expected that
a theory with two Debye energies will sometimes describe more subtle
effects. In addition, it becomes possible to apply such a model for
calculating crystals of various symmetries with the help of the
elastic moduli, if we make use of the method of the approximating
isotropic medium that was considered above.

\begin{figure}[h]
\vspace{0mm} \hspace{-0mm}
\includegraphics[width = 8.0cm]{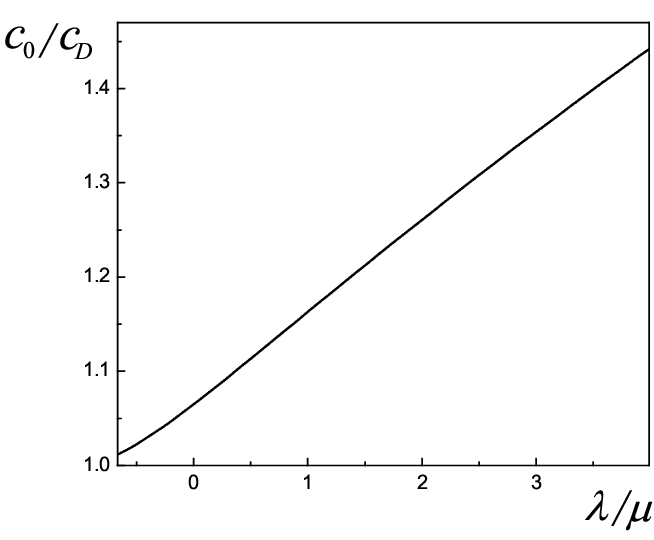} 
\vspace{-4mm}
\caption{\label{fig01} %
Dependence of the ratio of the phonon average velocities
$c_0\big/c_D$, defined by the formulas (\ref{31}) and (\ref{32}), on
the ratio of the Lame coefficients $\lambda/\mu$.
}%
\end{figure}

In this section, we formulate the Debye model without using the
averaging of the phonon velocity. It is natural to call such a model
a two-parameter one. The usual Debye model will also be called
one-parameter model. With an optimal choice of the phonon velocities
(\ref{28}), the free energy takes the form
\begin{equation} \label{34}
\begin{array}{l}
\displaystyle{%
  F=\frac{3N}{8}\big(\Theta_l+2\Theta_t\big)+\frac{NT}{3}\Big[3\ln\!\big(1-e^{-\beta\Theta_l}\big)-D\big(\beta\Theta_l\big)\Big]+ %
  \frac{2NT}{3}\Big[3\ln\!\big(1-e^{-\beta\Theta_t}\big)-D\big(\beta\Theta_t\big)\Big]. %
}
\end{array}
\end{equation}
The entropy $S=\partial F\big/\partial T$ and the energy $E=F+TS$
are given by the formulas:
\begin{equation} \label{35}
\begin{array}{l}
\displaystyle{%
  S=-\frac{N}{3}\Big[3\ln\!\big(1-e^{-\beta\Theta_l}\big)-4D\big(\beta\Theta_l\big)\Big] %
  -\frac{2N}{3}\Big[3\ln\!\big(1-e^{-\beta\Theta_t}\big)-4D\big(\beta\Theta_t\big)\Big], %
}
\end{array}
\end{equation}
\begin{equation} \label{36}
\begin{array}{l}
\displaystyle{%
  E=\frac{3N}{8}\big(\Theta_l+2\Theta_t\big)+NTD\big(\beta\Theta_l\big)+2NTD\big(\beta\Theta_t\big). %
}
\end{array}
\end{equation}
The heat capacity $C_V=T\big(\partial S\big/\partial T\big)$, with
taking into account (\ref{35}), is determined by the formula
\begin{equation} \label{37}
\begin{array}{l}
\displaystyle{%
  C_V=N\!\left[4D\big(\beta\Theta_l\big)-\frac{3\beta\Theta_l}{e^{\beta\Theta_l}-1}\right]+ %
      2N\!\left[4D\big(\beta\Theta_t\big)-\frac{3\beta\Theta_t}{e^{\beta\Theta_t}-1}\right]. %
}
\end{array}
\end{equation}
Here, the Debye energies are defined in the same way as above
$\Theta_t=\hbar c_t k_D$,\,$\Theta_l=\hbar c_l k_D$, but now the
velocities are expressed through the invariants of the elastic
moduli tensor of a crystal by the optimal relations (\ref{28}). If
in the formulas (\ref{34})\,--\,(\ref{37}) the Debye energies are
formally set equal, then we obtain the corresponding formulas of the
standard Debye theory. However, since the inequality
$c_l\big/c_t>\!\sqrt{4/3}$ holds in an isotropic medium, the same
inequality $\Theta_l\big/\Theta_t>\!\sqrt{4/3}$ holds as well for
the Debye energies, and therefore these energies cannot be equal.

Since it is more common to use one energy instead of two Debye
energies, let us introduce instead of the two parameters of the
energy dimension $\Theta_t,\,\Theta_l$ the single average Debye
energy $\Theta$ and one dimensionless parameter, for which it is
convenient to choose the angle $\chi$:
\begin{equation} \label{38}
\begin{array}{l}
\displaystyle{%
  \Theta^2=\frac{1}{3}\big(2\Theta_t^2+\Theta_l^2\big),\qquad \sin\chi=\sqrt{\frac{2}{3}}\frac{\Theta_t}{\Theta},\qquad %
  \cos\chi=\frac{1}{\sqrt{3}}\frac{\Theta_l}{\Theta},\qquad {\rm tg}\,\chi=\sqrt{2}\,\frac{\Theta_t}{\Theta_l}. %
}
\end{array}
\end{equation}
The average Debye energy $\Theta$ determined in such a way
corresponds to the averaging of velocities (\ref{32}). Owing to the
given above inequality $\Theta_l\big/\Theta_t>\!\sqrt{4/3}$\, for
the ratio of the Debye energies, the tangent of the angle $\chi$
introduced in (\ref{38}) takes the maximum value at the angle
$\chi_m$, which is determined by the condition ${\rm
tg}\,\chi_m=\sqrt{3/2}$, whence $\chi_m=0.886$\, and, accordingly,
$\cos\chi_m=\sqrt{2/5}$, $\sin\chi_m=\sqrt{3/5}$. Thus, the angle
can vary within the range $0\leq\chi<\chi_m$. Here are also given
the expressions for the parameters defined in (\ref{38}) in terms of
the invariants of the crystal elasticity tensor:
\begin{equation} \label{39}
\begin{array}{l}
\displaystyle{%
  \Theta^2=\frac{\hbar^2k_D^2}{9\rho}\lambda_B,\qquad \sin^2\chi=\frac{\big(3\lambda_B-\lambda_A\big)}{5\lambda_B},\qquad %
  \cos^2\chi=\frac{\big(2\lambda_B+\lambda_A\big)}{5\lambda_B}. %
}
\end{array}
\end{equation}
Let us consider in more detail the behavior of the heat capacity
(\ref{37}) with temperature. As is known, in the standard Debye
theory the heat capacity is a universal function of the ratio
$T\big/\Theta_D$ \cite{LL2}. According to this model the heat
capacities of all bodies being in corresponding states, i.e. having
the same $T\big/\Theta_D$, should be the same. For real bodies this
law, obviously, is not satisfied. In the proposed two-parameter
model, as we see, the law of the corresponding states does not hold,
since the heat capacity also depends on the parameter $\chi$ which
is different for different bodies (\ref{39}). At high temperatures
$T\gg\Theta$, the formula for the heat capacity has visually the
same form as in the usual theory
\begin{equation} \label{40}
\begin{array}{l}
\displaystyle{%
  C_V=3N\!\left[1-\frac{1}{20}\left(\frac{\Theta}{T}\right)^2\right]. %
}
\end{array}
\end{equation}
Here, the average Debye energy is determined by the relations
(\ref{38}), (\ref{39}) and the dependence on the parameter  $\chi$
falls out in this approximation. At low temperatures $T\ll \Theta$,
the same as in the usual theory, the cubic dependence on temperature
is preserved
\begin{equation} \label{41}
\begin{array}{l}
\displaystyle{%
  C_V=\frac{12}{5}\pi^4Nf(\chi)\left(\frac{T}{\Theta}\right)^3, %
}
\end{array}
\end{equation}
but in this case the coefficient also depends on the parameter
$\chi$ through the function
\begin{equation} \label{42}
\begin{array}{l}
\displaystyle{%
  f(\chi)\equiv\frac{1}{3^{5/2}}\left(\frac{1}{\cos^3\chi}+\frac{2^{5/2}}{\sin^3\chi}\right) %
}
\end{array}
\end{equation}
and turns out to be different for different substances. The
influence of the second parameter $\chi$ is manifested most
significantly at low temperatures. At high temperatures, the main
difference of the two-parameter theory consists in that the average
Debye energy (\ref{38}), (\ref{39}) is defined differently here and,
moreover, it is related to the invariants of the elasticity tensor
of a crystal of a certain symmetry (\ref{39}).

\begin{figure}[b!]
\vspace{2mm} \hspace{-0mm}
\includegraphics[width = 8.0cm]{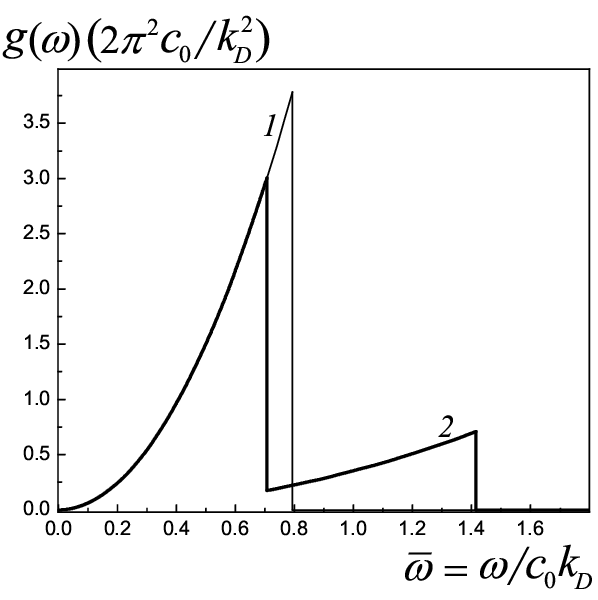} 
\vspace{-4mm}
\caption{\label{fig02} %
The densities of phonon states
$g(\omega)\big(2\pi^2c_0\big/k_D^2\big)$ as functions of the
frequency $\overline{\omega}=\omega\big/c_0k_D$; {\it 1} -- the
one-parameter model, {\it 2} -- the two-parameter model. The
calculation is made for $\lambda/\mu=2$.
}%
\end{figure}

Let us also give an expression for the density of phonon states,
which in this case is determined by the formula
\begin{equation} \label{43}
\begin{array}{l}
\displaystyle{%
  g(\omega)= \int\!\!\frac{d{\bf k}}{(2\pi)^3}\Big[\delta\big(\omega-c_lk\big)+2\delta\big(\omega-c_tk\big)\Big]. %
}
\end{array}
\end{equation}
As a result, we have
\begin{equation} \label{44}
\begin{array}{l}
\displaystyle{%
  g(\omega)\,2\pi^2= \left\{
               \begin{array}{l}
                 \omega^2\left(1\big/c_l^3+2\big/c_t^3\right),\quad \omega< k_Dc_t, \vspace{2mm} \\ %
                 \omega^2\big/c_l^3, \hspace{20mm} k_Dc_t<\omega<k_Dc_l, \vspace{2mm} \\ %
                 0, \hspace{28mm}                  \omega>k_Dc_l. %
               \end{array} \right. %
}
\end{array}
\end{equation}
The form of the density of phonon states is shown in Fig.\,2. Due to
the presence of the two phonon branches in the function $g(\omega)$,
a jump appears at the frequency $\omega=k_Dc_t$ and, therefore, the
structure of the density of states proves to be more complicated
than in the standard model. As is known \cite{Kittel,AM}, there are
van Hove features in the density of phonon states of real crystals,
so that the two-parameter model is closer in this respect to real
crystals.

\section{CRYSTALS OF THE CUBIC SYSTEM}\vspace{-0mm} %
As an example, we consider the application of the two-parameter
Debye theory to crystals of the cubic system, which are most close
to an isotropic medium and are characterized by three elastic
moduli. In this case the two invariants are given by the expressions
$\lambda_A=3c_{11}+6c_{12}$, $\lambda_B=3c_{11}+6c_{44}$ in terms of
three elastic moduli of the cubic crystal in the standard notation
\cite{Fedorov}, and the approximating Lame coefficients are:
\begin{equation} \label{45}
\begin{array}{l}
\displaystyle{%
  \tilde{\lambda}=\frac{1}{5}\big(c_{11}+4c_{12}-2c_{44}\big),\qquad  %
  \tilde{\mu}=\frac{1}{5}\big(c_{11}-c_{12}+3c_{44}\big).  %
}%
\end{array}
\end{equation}
The values of these quantities for some cubic crystals at low
temperatures are shown in Table 1.
\begin{table}[h!] \nonumber
\vspace{-4mm}
\centering %
\caption{The values of the adiabatic elastic moduli of tungsten,
copper and lead at low temperatures \cite{Kittel}, the invariants
$\lambda_A,\lambda_B$ and the approximating Lame coefficients
(\ref{45}) (in units of $10^{12}\,dyn/cm$).} %
\vspace{0.5mm}%
\begin{tabular}{|c|c|c|c|c|c|c|c|} \hline 
Crystal  &   $c_{11}$       &   $c_{12}$      &  $c_{44}$       &  $\lambda_A$ &  $\lambda_B$ &$\tilde{\lambda}$&  $\tilde{\mu}$  \\ \hline %
W        &\,\,\, 5.326\,\,\,&\,\,\,2.049\,\,\,&\,\,\,1.631\,\,\,&\,\,28.272\,\,&\,\,25.764\,\,&\,\,\,2.052\,\,\,&\,\,\,1.634\,\,\,\\ \hline %
Cu       &  1.762           &      1.249      &      0.818      &    12.78     &    10.194    &      1.024      &      0.593      \\ \hline %
Pb       &  0.555           &      0.454      &      0.194      &    4.389     &    2.829     &      0.397      &      0.137      \\ \hline %
\end{tabular}  
\end{table}

\noindent Table 2 shows the density, the transverse and longitudinal
phonon velocities calculated using the data of Table 1
and the average velocities determined by the formulas (\ref{31}) and (\ref{32}). %
\begin{table}[h!] \nonumber
\vspace{-4mm}
\centering %
\caption{The values of the density $\rho$ ($g/cm^3$), transverse
$c_t$ and longitudinal $c_l$ velocities (\ref{28}) and also average
velocities $c_{D}$ (\ref{31}) and $c_{0}$ (\ref{32}) (in units of
$10^{6}\,cm/s$).} %
\vspace{0.5mm}%
\begin{tabular}{|c|c|c|c|c|c|} \hline 
Crystal  &   $\rho$        &      $c_{t}$    &     $c_{l}$     &      $c_{D}$    &      $c_{0}$    \\ \hline %
W        &\,\,\,1.551\,\,\,&\,\,\,0.292\,\,\,&\,\,\,0.524\,\,\,&\,\,\,0.225\,\,\,&\,\,\,0.385\,\,\,\\ \hline %
Cu       &      9.018      &      0.257      &      0.495      &      0.199      &       0.355      \\ \hline %
Pb       &      11.60      &      0.109      &      0.240      &      0.085      &       0.165      \\ \hline %
\end{tabular}  
\end{table}

\noindent Table 3 gives the Debye wave number $k_D$, parameter
$\chi$ (\ref{38}), value of the function $f(\chi)$ (\ref{42}) and
Debye temperatures calculated by the formulas $\Theta_t=\hbar
c_tk_D$, $\Theta_l=\hbar c_lk_D$, $\Theta_D=\hbar c_Dk_D$,
$\Theta\equiv\Theta_0=\hbar c_0k_D$, where the velocity values are
taken from Table 2, as well as the experimental low-temperature
value of the Debye temperature $\Theta_{D\rm exp}$ \cite{Kittel}.
\begin{table}[h!] \nonumber
\vspace{-4mm}
\centering %
\caption{The wave number $k_D$ (in $10^{8}\,cm^{-1}$), Debye temperatures in Kelvin degrees ($K$).} %
\vspace{0.5mm}%
\begin{tabular}{|c|c|c|c|c|c|c|c|c|} \hline 
Crystal  &       $k_D$     &     $\Theta_t$        &     $\Theta_l$    &$\Theta=\Theta_0$  &  $\Theta_D$       &$\Theta_{D\rm exp}$&      $\chi$     &      $f(\chi)$  \\ \hline %
W        &\,\,\,1.551\,\,\,&\,\,\,\,\,346\,\,\,\,\,&\,\,\,\,\,621\,\,\,\,\,&\,\,\,\,456\,\,\,\,&\,\,\,\,\,266\,\,\,\,\,&\,\,\,\,\,400\,\,\,\,\,&\,\,\,0.665\,\,\,&\,\,\,1.676\,\,\,\\ \hline %
Cu       &      1.717      &        337            &          650          &      466          &          261          &          343          &      0.632      &      1.882      \\ \hline %
Pb       &      1.263      &        105            &          232          &      159          &           82          &          105          &      0.569      &      2.427      \\ \hline %
\end{tabular}  
\end{table}

As we can see, the Debye temperatures $\Theta_D$ calculated in the
standard one-parameter model from the values of elastic constants
prove to be much lower than the measured Debye temperatures
$\Theta_{D\rm exp}$. The calculation of the Debye temperatures in
the proposed two-parameter model from the values of elastic
constants gives values greater than $\Theta_{D\rm exp}$. Introducing
to the theory the second parameter most appreciably affects the
calculation of thermodynamic quantities at low temperatures. Figure
3 shows the calculated temperature dependencies of the heat capacity
of copper at low temperatures. The calculations are performed using
the approximated values of the elastic moduli for the one-parameter
(curve \textit{1}) and two-parameter (curve \textit{2}) models.
Curve 3 is plotted using the experimental value of the Debye
temperature. The relative error $\delta=\big(C_V-C_{V\rm
exp}\big)\big/C_{V\rm exp}$ in the one-parameter model for copper is
$\delta_1=1.27$. In the two-parameter model the discrepancy with the
experimental value proves to be much smaller: $\delta_2=0.25$. The
values of these deviations for tungsten are $\delta_1=2.7$,
$\delta_2=0.57$, and for lead $\delta_1=1.1$, $\delta_2=0.3$. As
seen, the calculation in the one-parameter model gives a greater
discrepancy with experimental data. At the same time, the
two-parameter model leads to much better agreement with experiment
and can be used for theoretical estimates of the Debye temperature
from the values of the crystal elastic moduli.
\begin{figure}[h]
\vspace{-6mm}\hspace{-0mm}
\includegraphics[width = 8.0cm]{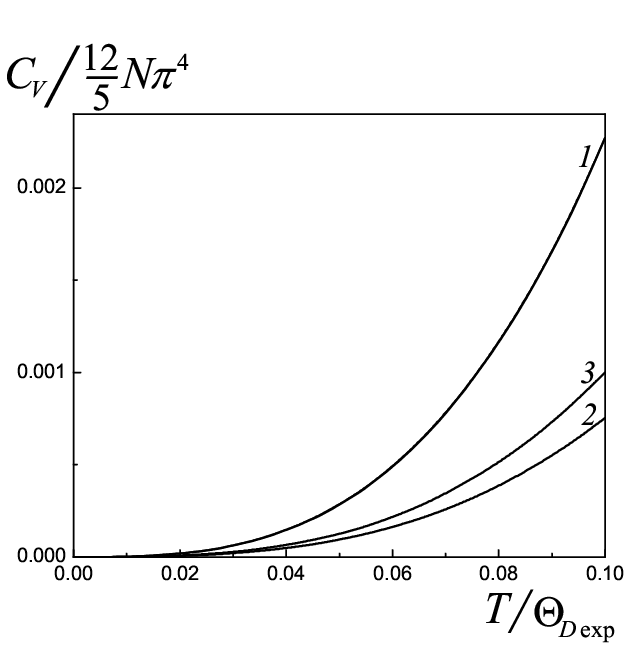} 
\vspace{-4mm}
\caption{\label{fig03} %
The temperature dependence of the heat capacity of copper at low
temperatures: {\it 1} -- the calculation through the approximated
elastic moduli in the one-parameter model, {\it 2} -- the similar
calculation in the two-parameter model, {\it 3} -- the dependence in
the standard model, where the Debye temperature is considered as a
phenomenological parameter. The Debye temperature for Cu
$\Theta_{D\rm exp}=343\,K$.
}%
\end{figure}

\vspace{-3mm}
\section{Conclusions}\vspace{-2mm} %

It is shown that the proposed earlier method \cite{Fedorov} of
describing the elastic properties of crystals on the basis of a
comparison with an isotropic medium follows from the requirement of
the maximal closeness of the free energies of a crystal and an
isotropic medium. In this work it is proposed a model of an
isotropic elastic medium which is similar to the standard Debye
model and in which the existence of both transverse and longitudinal
phonons is taken into account. In this model, besides the Debye
energy, an additional parameter is introduced, so that the law of
corresponding states characteristic of the usual Debye theory ceases
to be fulfilled. The calculation of the heat capacity at low
temperatures using the approximated elastic moduli leads to much
better agreement with experimental data than for the usual theory
using the average phonon velocity.

There are two extreme points of view on the Debye model. Often it is
given an unduly general meaning to the Debye model and, when
processing experimental data, observable quantities are adjusted to
the relations of this theory assuming that the Debye energy depends
on temperature. The opposite point of view consists in that the
relations of the Debye theory are considered as crude interpolation
formulas \cite{LL,AM}. The Debye model, of course, is a quite
approximate and simple (in what its value consists) model of the
solid body, but, in our opinion, its value is not limited only to
the possibility of constructing a single interpolation formula that
would correctly describe the behavior of a body in the limit of low
and high temperatures. This model, as shown above, allows further
development and, in particular, generalization with taking into
account the interaction of phonons \cite{Poluektov,Poluektov2}, and
also can be extended to describe surface phenomena in solids.

\end{document}